# The Impact of Grid Storage on Balancing Costs and Carbon Emissions in Great Britain

Seyyed Mostafa Nosratabadi[1,5,*], Iacopo Savelli[2], Volkan Kumtepeli[1], Phil Grunewald[1], Marko Aunedi[3], David A. Howey[1,4], Thomas Morstyn[1]

[1]Department of Engineering Science, University of Oxford, OX1 3PJ, Oxford, UK

[2]GREEN Research Centre, Bocconi University, Milan, 20136, Italy

[3]Department of Electronic and Electrical Engineering, Brunel University London, Uxbridge UB8 3PH, UK

[4]Faraday Institution, Quad One, Harwell Campus, Didcot, OX11 ORA, UK

[5]Lead contact

*Correspondence: mostafa.nosratabadi@eng.ox.ac.uk, sm.nosratabadi@yahoo.com

## SUMMARY

Grid energy storage can help to balance supply and demand, but its financial viability and operational carbon emissions impact is poorly understood because of the complexity of grid constraints and market outcomes. We analyse the impact of several technologies (Li-ion and flow batteries, pumped hydro, hydrogen) on Great Britain's balancing mechanism, the main market for supply-demand balancing and congestion management. We find that, for many locations and technologies, financially optimal operation of storage for balancing can result in higher carbon emissions. For example, the extra emissions associated with a 1 MW 2-hour duration Li-ion battery in winter vary between +230 to -71 kgCO2/h. Although storage enable higher usage of renewables, it can also unlock additional demand leading to greater use of gas. In addition, balancing services alone are presently insufficient for financial viability of storage projects. This work highlights the need for market reform aligning financial incentives with environmental impacts.

## INTRODUCTION

The energy sector is undergoing a significant transformation worldwide as countries move towards cleaner power generation.[1] New solutions for clean flexibility are required to manage network power flows and balance supply and demand of electricity in real-time as renewable energy generation increases.[2] Energy storage has emerged as a promising option, and the IEA projects largescale global rollout of storage in the coming years.[3] There has been significant research investigating the optimal deployment of grid energy storage as part of the power system decarbonisation. Previous works have looked at different geographies[4-6], and the suitability of different energy storage technologies for providing specific grid services.[7] Missing from this research, however, are detailed studies of real markets to explore whether investment in storage is actually incentivised. The business case for storage is clear for certain services, such as frequency regulation.[8,9] However, analysing the business case for other applications such as system balancing is challenging, requiring detailed understanding of market behaviours and grid power flows.[10]

Additionally, it is not clear that the current financial incentives for developing and operating grid storage align with the environmental goal of reducing carbon emissions. It has been shown that grid storage may decrease emissions in cleaner grids dominated by nuclear and renewables, but there is also a risk that storage may increase emissions in some situations, by shifting the generation mix towards dirtier sources.[5] There has been detailed work on the full lifecycle emissions of individual storage systems[11], but modelling the systems-level carbon emissions impact is a more recent research topic. The existing literature tends to consider simplified scenarios, either using national marginal emissions factors[12], or economic dispatch ignoring transmission grid physical constraints and losses.[13] There is a lack of simulation models and tools that can provide the granularity necessary to understand the carbon emissions impact associated with storage location, operation and technology choice.[14]

The Great Britain (GB) power system, which supplies England, Scotland, and Wales combined, is an interesting case study for investigating the market incentives for grid storage because it has seen rapidly growing large-scale investment in renewable energy and storage over the past decade and has ambitious targets to achieve net-zero greenhouse gas emissions electricity by 2030.[15,16] Renewable power generation now supplies 47.3% of the GB annual electricity demand[17], and battery storages provides around 15% of GB balancing service.[18] However, GB still relies heavily on gas generation for system balancing. This has motivated interest in energy storage as an alternative source of clean flexibility. It is projected that substantially more energy storage will be required to meet the UK's net-zero targets.[19] The GB region has used pumped hydro storage for decades, but in the past five years, battery energy storage build-out has rapidly accelerated, from 0.88 GWh to 4.6 GWh operational.[20,21]

The business case for grid storage for GB frequency regulation is well established but the market size is limited.[18] There is a larger market opportunity if storage can provide services in the balancing mechanism (BM), which is the primary tool of the GB National Grid Electricity System Operator (NGESO) to balance supply and demand and manage transmission constraints. The NGESO operates the BM, procuring the additional power required for system balancing.[22,23] Detailed analysis of the BM has shown that network location strongly influences the impact of a generation unit on overall system cost and carbon emissions, due to how units are re-dispatched to address transmission constraints.[24]

Critical unresolved issues regarding the integration of grid storage are, firstly, how can the integration of storage improve the cost of system balancing, secondly, to what extent do existing market incentives align efficient system balancing with carbon emissions reduction, and finally, how much does location influence the system value and financial viability of storage.

To address these topics, we developed a high-fidelity transmission system and BM model and simulated the integration of different storage technologies and their impact on the BM in the GB context. We considered lithium-ion batteries (LIBs), vanadium redox flow batteries (VRFBs), pumped storage hydropower (PSH), and hydrogen energy storage (HES), evaluating the benefits and challenges in BM cost and carbon emissions reduction associated with each technology, and their locational effectiveness within the GB power grid. In essence, we focus on determining the degree to which BM cost and the optimal dispatch of storage are in harmony with the carbon emissions reduction objective. We find that, in many cases, operating storage purely for financial benefit for system balancing results in *higher* overall carbon emissions. Results vary considerably depending on geographic location within GB, season, and storage technology properties. There is currently no incentive for storage developers to place projects in the locations where they can have the largest impact on grid carbon emissions. This has important implications regarding the need for new market mechanisms on the pathway to net-zero, such as including stronger carbon prices, or incorporating explicit carbon emissions goals into the BM dispatch procedure.

## RESULTS

Four different storage technologies were investigated, both at a marginal size (1 MW) and at a larger scale (100 MW), spanning various geographic locations. We compared the system operation in winter and summer, with and without grid storage included. We report the impact of storage on balancing costs and carbon emissions, and an analysis of the net present value of storage.

### Model and dataset

Simulations of the GB transmission grid and balancing mechanism were conducted using a high-fidelity transmission system model coupled with an optimisation approach emulating the redispatch of thousands of individual balancing units. The transmission system model includes all 1982 nodes and transmission lines at 33 kV and above. Demand data and market data (bids, offers and outcomes) for all balancing units above 1 MW was included. Reported demands in each geographic area were linked to transmission grid supply points using the explained approach in section S.3 of supplemental information. Overall, full details of all models and assumptions are given in supplemental information.

### Marginal impact of different energy storage systems employment in winter and summer seasons

This section reports the marginal impact of employing the described energy storage systems, focusing on their effects on the BM cost and carbon emissions. In addition, the impact evaluation of energy storage system integration on various generation levels is also performed in section S.4 of supplemental information.

*Marginal impact on BM cost*

This subsection is focused on the marginal impact of energy storage systems integration on BM cost. For this, our GB transmission system and balancing mechanism model has been implemented and solved in the presence of 1 MW (2-hour duration) energy storage system integration. The optimal BM cost values have been calculated for the various energy storage system technologies located in different GSP groups across the GB network. These values are visually represented as color-coded maps in Figure 1, showcasing the marginal impact on BM cost in comparison to the base case (without storage system) for both January and July 2022. The numerical results are drawn from Tables S1 and S2 (see supplemental information). The obtained results underscore that the integration of these technologies leads to a noteworthy reduction in BM costs, particularly during the summer season. In Figure 1, first four maps, we observe that for all four energy storage system types have the potential to decrease BM costs across all GSP groups in January. However, the extent of these savings varies based on the specific technology and the GSP group under consideration.

Notably, the South and North of Scotland GSP groups stand out as having the highest potential for cost savings from energy storage system integration. This can be attributed to the substantial share of renewable energy generation in these regions, which tends to be intermittent during the winter months. The energy storage systems mitigate the variability associated with renewable generation and reducing the import of electricity from other GSP groups. Conversely, the integration of HES technology within some GSP groups appears to have a comparatively lower impact on BM costs. This is due to low round-trip efficiency which means HES is rarely dispatched.

During the summer season, we encounter higher BM cost reductions in most of the GSP groups. The best value in the summer month is approximately double that of the best value observed in the winter month. It can be seen that LIB and PSH integration within the H GSP group can achieve the most reduction in monthly BM cost (~£3,800 to ~£4,000 for 1 MW storage integration). Due to its low efficiency, when HES is integrated in certain GSP groups, specifically A, B, K, and L, there is no change in operation, resulting in a cost difference of zero. However, HES integration still achieves a reasonable monthly BM cost reduction in the N and P GSP groups in the summer season (~£3,000 for 1 MW storage integration).

*Marginal impact on carbon emission*

An essential component of our study is the impact of energy storage systems on carbon emissions. Through the incorporation of 1 MW storage units (2-hour duration) into different GSP groups within the grid, and utilizing the carbon intensity data, we have generated the maps presented in Figure 2. These maps show the emission differences with respect to the base case (without the storage system), as detailed in Tables S3 and S4 (see supplemental information) for the winter and summer seasons, respectively. In Figures 2, first four maps, we explore the winter season results for the four storage technologies: LIB, VRFB, PSH, and HES. The most favourable outcomes are observed in GSP groups E, G, B, and P for LIB, VRFB, PSH, and HES integration, respectively. Comparing these results alongside the cost-related findings in the first four maps of Figure 1, it can be seen that reduction within a specific GSP group does not invariably translate into a reduction in emissions.

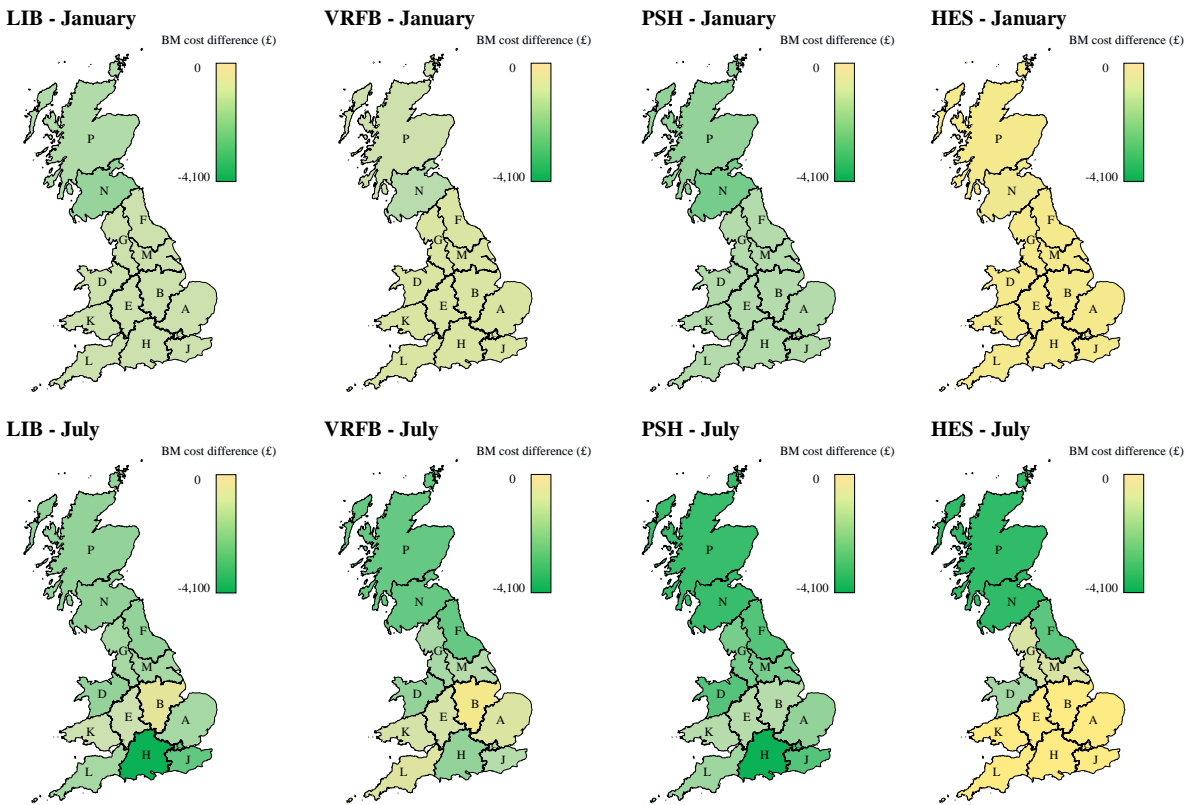

**Figure 1: The marginal impacts of storages on cost vary strongly with location and technology**

The GB map shows the monthly BM cost difference of 1 MW storage integration case w.r.t the base case in different GSP groups (values are reported in Tables S1 and S2). The first four maps show the results for January 2022 and the rest show the results for July 2022.

Furthermore, the integration of storage systems may trigger multiple changes in accepted BM unit bids and offers, each with a different carbon intensity profile. This dynamic may lead to an increase or decrease in overall emissions.

For the summer season, the results are depicted in the second four maps of Figure 2. Carbon emissions have been reduced slightly in some GSP groups (in 10, 10, 8, and 5 GSP groups respectively for LIB, PSH, HES, and VRFB). LIB integration within most of the regions offers greater emissions reduction compared to the other technologies. Similar to the winter season, GSP groups achieving the most significant reductions in energy costs do not necessarily correlate with the highest reductions in carbon emissions.

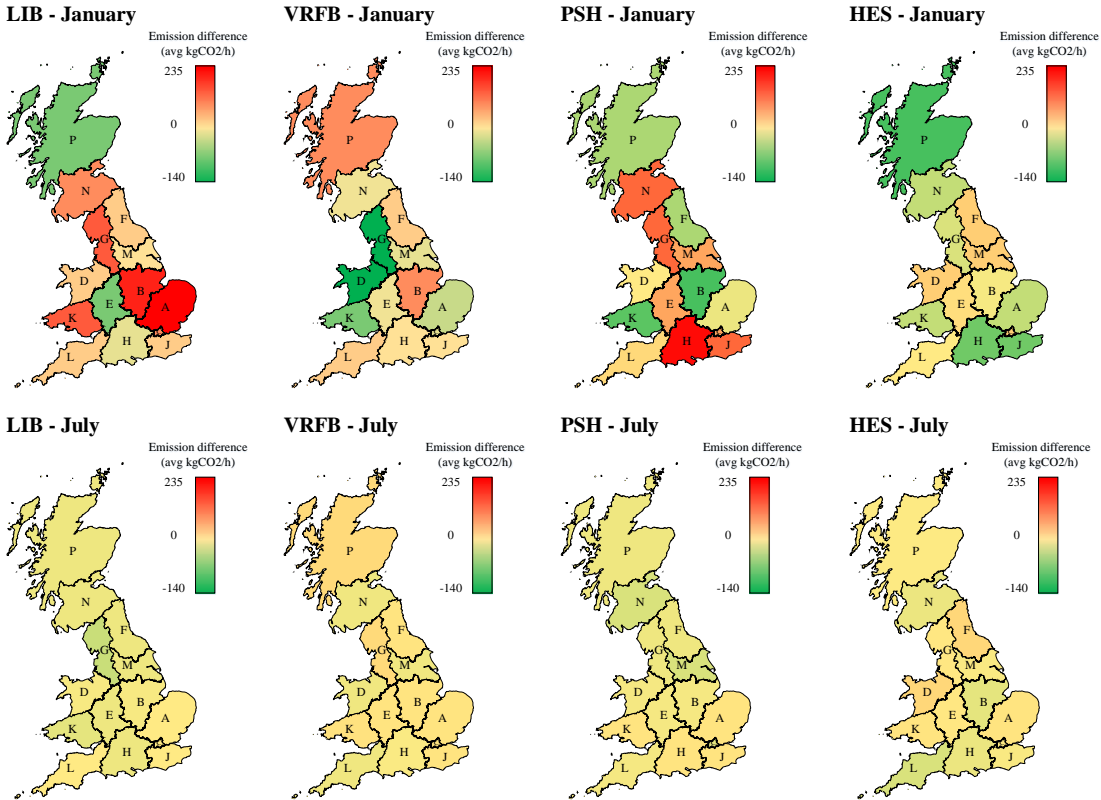

**Figure 2: The marginal impacts of storages on carbon emission vary strongly with location and technology**

The GB map shows the emission difference of 1 MW storage integration case w.r.t the base case in different GSP groups (values are reported in Tables S3 and S4). The first four maps show the results for January 2022 and the rest show the results for July 2022.

Comparing the winter and summer results for VRFB, HES, PSH, and LIB, we observe emission reductions in 8, 8, 5, and 3 GSP groups during winter, respectively (out of 14 total). In contrast, these figures have changed to 5, 8, 10, and 10 GSP groups with emission reductions during summer. If we consider the average of these figures as a number for whole year, it is concluded that HES can relatively show reduction in more GSP groups comparing to others.

**Grid-scale integration effect of different energy storage systems in winter and summer seasons**

This subsection explores the effect of integrating grid-scale storage systems (100 MW units) on operation during winter and summer. We aim to understand how adding a large storage unit in different locations affects BM cost and carbon emissions. As with the marginal assessment, the effect on energy generation mix is also investigated in section S.4 of supplemental information.

*Grid-scale integration effect on BM cost*

We present the BM cost difference results comparing the results with and without 100 MW storage (2-hour duration) added, depicted through color-coded maps in Figure 3. Segmented into first four maps of Figure 3, the visualizations illustrate the results in January for different storage technologies. In all cases, the integration of storage systems yields a reduction in BM costs. Significantly, the N GSP group emerges as the most favourable GSP group for integration in which the monthly BM cost reduction range from ~£15,500 with HES to ~£166,500 with PSH. Conversely, the regions denoted as F, B, F, and P respectively exhibit the lowest monthly BM cost reductions for LIB and PSH, VRFB, and HES integration so that the lowest value is from HES integration in P GSP group at ~£10,900. The lower BM cost reduction in these regions can be attributed to either bid/offer values of different energy sources within the GSP group or the low efficiency of the integrated storage technology.

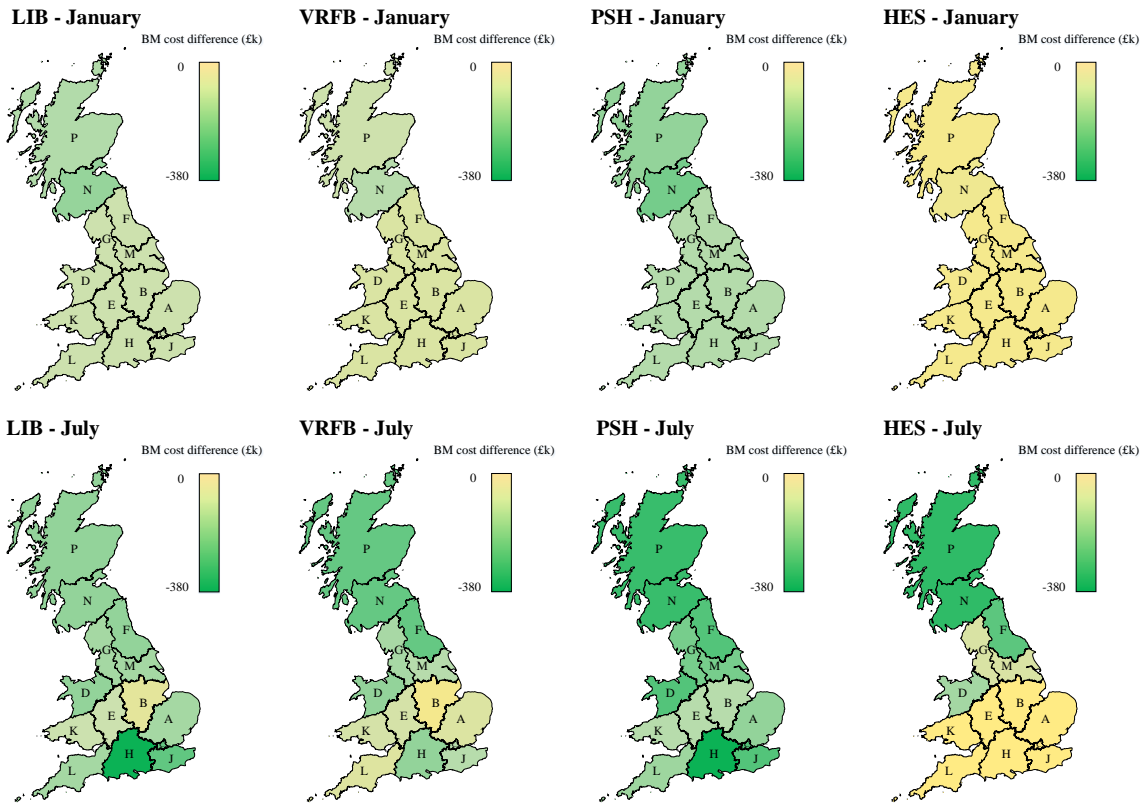

**Figure 3: The impacts of grid-scale storages on cost vary strongly with location and technology**

The GB map shows the monthly BM cost difference of 100 MW storage integration case w.r.t the base case in different GSP groups (values are reported in Tables S5 and S6). The first four maps show the results for January 2022 and the rest show the results for July 2022.

Moving on to the second four maps of Figure 3, we shift our focus to the maps representing the BM cost differences in July, specifically pertaining to LIB, VRFB, PSH, and HES integration within GSP groups. The results for LIB and PSH integration exhibit a degree of similarity, likely attributed to their similar efficiencies i.e. 85% and 79%, respectively. A noteworthy observation emerges when considering the integration of HES, where four distinct GSP groups—A, B, K, and L—reveal no discernible BM cost difference. In stark contrast, the integration of PSH in the H GSP group yields the most substantial monthly cost difference, amounting to ~£372k, surpassing all other storage integration results.

*Grid-scale integration effect on carbon emission*

The emission difference maps for 100 MW (2-hour duration) integration are presented in Figure 4. First four maps of Figure 4 specifically pertaining to the integration of different energy storage systems during January 2022. GSP group B stands out in emission reduction for LIB integration. LIB in this region gives the greatest emissions reduction (~216 kgCO2/h) across all storage types and locations. LIB, VRFB, HES and PSH integration emissions reductions only occur in 4, 1, 1, and 0 GSP group(s), respectively. Second four maps of Figure 4 provide a visual representation of the carbon emission differences resulting from the integration of different energy storage systems within the various GSP groups for July 2022. Likewise, in these cases, the observations reveal that the number of GSP groups exhibiting significant emission reduction is relatively limited. Additionally, the integration of PSH indicates an increase in carbon emissions of all GSP groups. The best case is the integration of LIB within the F GSP group while GSP groups D, F, N, and P show the worst results with HES integration. VRFB increases carbon emissions with respect to the base case when integrated in all GSP groups except GSP group M.

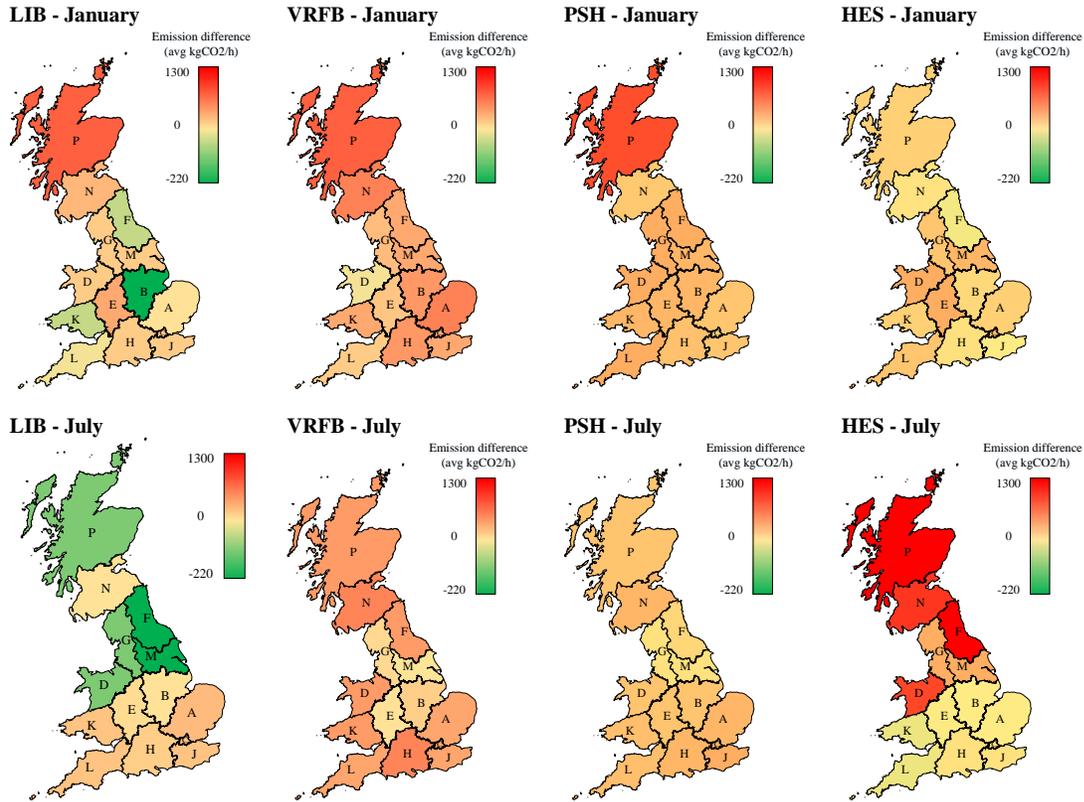

**Figure 4: The impacts of grid-scale storages on carbon emission vary strongly with location and technology**

The GB map shows the emission difference of 100 MW storage integration case w.r.t the base case in different GSP groups (values are reported in Tables S7 and S8). The first four maps show the results for January 2022 and the rest show the results for July 2022.

### NPV assessment and sensitivity analysis

In this section, we assess the NPV associated with the deployment of various energy storage systems. Here, value stacking of BM participation along with the Capacity Market (CM) and Dynamic Containment (DC) service are considered. DC is GB's fast acting post-fault frequency regulation service. The storage capacity percentage for participation in DC, which varies between 39% to 54% depending on technology, location, and season, corresponds to the remaining capacity after BM participation. CM participation is allowed on top of BM and DC participation. Given the interplay between these services, it is essential to understand the distinct and combined roles of storage system in each market to optimize overall system performance and economic returns.

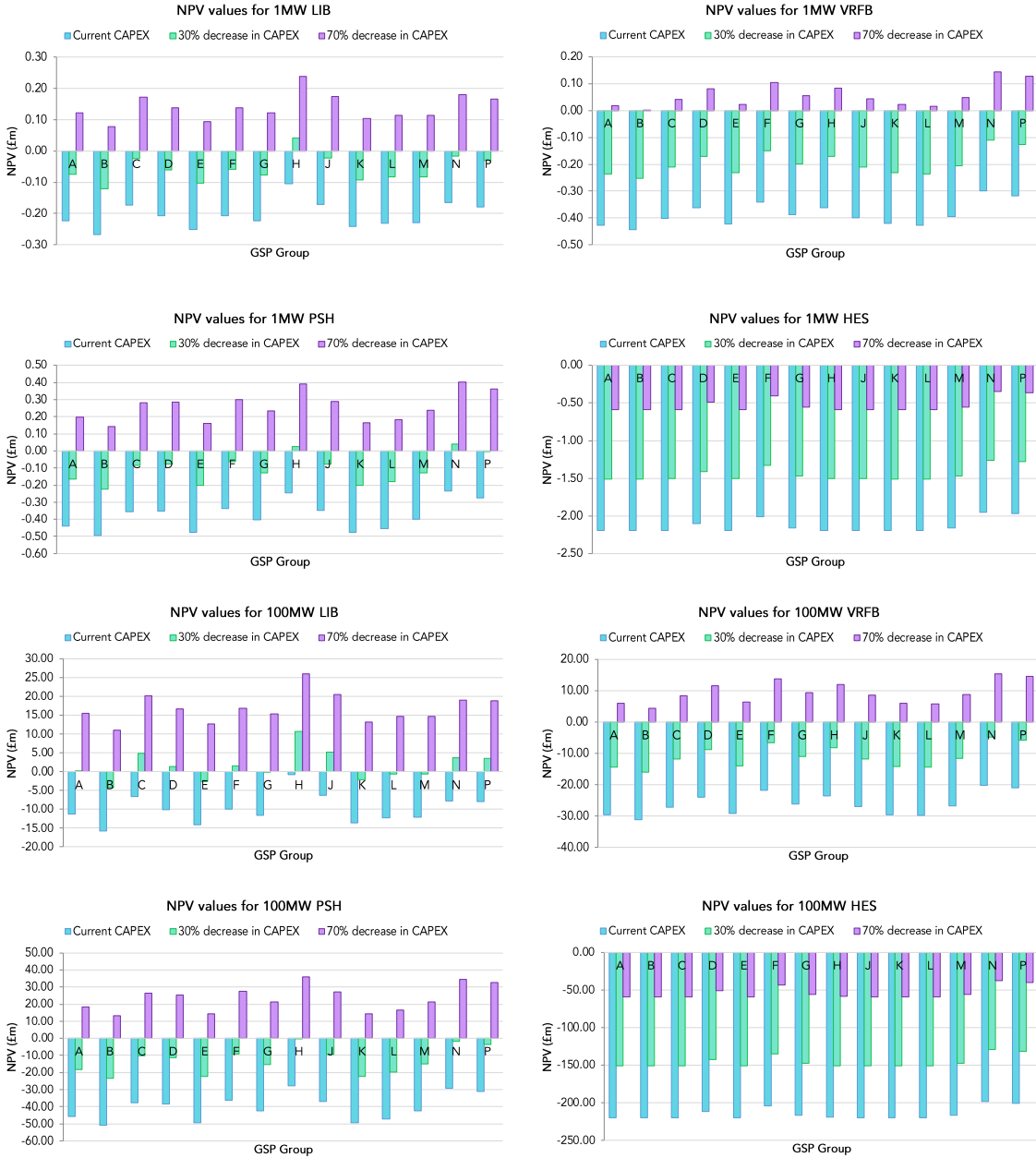

**Figure 5: The storage technology profitability varies based on revenue obtained in different locations and costs imposed**

Charts show NPV values for different 1 MW and 100 MW storage systems integrated into various GSP groups within the GB network by current CAPEX, by 30% decrease in CAPEX, and by 70% decrease in CAPEX.

We are particularly interested in the value of energy storage system in the BM since the BM is the main market mechanism used for balancing supply and demand and managing transmission constraints within the GB network. Figure 5 shows the NPV values for the different storage technologies for 1 MW and 100 MW (2-hour duration) integration. First four charts of Figure 5 show the NPV results for 1 MW storage integration. At current *CAPEX* costs the technologies are not economically viable. LIB gives the least negative NPV values, followed closely by PSH. From a locational point of view, GSP Groups H

(Southern) and N (South of Scotland) give higher NPV results for LIB and PSH integration, respectively, while GSP group N (South of Scotland) is the best for VRFB and HES integration.

Turning our attention to the 100 MW integration case, second four charts of Figure 5 provide the NPV values associated with storage system integration across various GSP groups. In the fifth chart of Figure 5, the NPV values for LIB are consistently negative across all GSP groups, indicating a financial deficit. The best GSP group is H, where a relatively small reduction in *CAPEX* could potentially render LIB profitable. The NPV values for the HES, as illustrated in the last chart of Figure 5, are very low given current *CAPEX*.

To demonstrate how removing DC revenue stacking would worsen the outcome, we can examine the NPV results derived solely from BM and CM revenues. For example, if we focus on NPV value for 100 MW case of LIB in GSP group H, the value would be ~-£9m while it would be improved to ~-£800k by considering revenue from DC. This shows the importance of including DC revenue in the stacking strategy for energy storage systems from a profitability perspective.

Sensitivity analysis shows the impact of energy storage system investment cost reductions on the overall NPV results. Specifically, we examine the effects of a 30% and 70% decrease in *CAPEX*. The effect of lower investment cost of energy storage systems on NPV is shown in Figure 5. It is evident that, for 1 MW storage integration, all four types of technologies yield better NPV values. For 30% *CAPEX* reduction, only LIB technology in GSP group H shows positive NPV. For 70% *CAPEX* reduction, the NPV values are positive for all technologies except HES. In the case of 100 MW LIB integration, 8 GSP groups show profitability under a 30% *CAPEX* reduction. The highest NPV is £10.68m for GSP group H. In this situation, the more number of GSP groups with positive NPV values by 100 MW case against 1 MW one is related to effectiveness of higher value of revenues via BM, CM and DC for 100 MW case comparing to 1 MW one. For 70% *CAPEX* reduction, NPV is positive for all GSP groups and again the NPV is highest at £26m for GSP group H. For 100 MW VRFB and PSH integration with 30% *CAPEX* reduction, none of the GSP groups show positive values but under the 70% *CAPEX* reduction scenario both show profitability with the highest values at £15.42m and £35.94m for GSP groups N and H, respectively. For HES integration all scenarios and GSP groups shown in the last chart of Figure 5 have negative NPV.

Overall, the NPV assessment and sensitivity analysis underscore that LIB and PSH deliver favourable results across all GSP groups compared to VRFB and HES. In the case of HES integration, all GSP groups consistently show negative NPV and this remains the case even for significant decreases in *CAPEX*. However, it should be noted that HES has not yet been widely deployed as a grid-scale storage technology, so there is uncertainty around how its cost and operational parameters like round trip efficiency may improve in future.

## DISCUSSION

This study has investigated the operation of prominent energy storage system technologies in the GB BM, with a particular focus on the impact of network location on cost and carbon emissions. Key findings can be summarized as follows:

**Impact of storage technology:** In this paper, NPV economic evaluation shows that if the investment cost of specific storage technologies can be reduced by 30% or more, integration can be viable economically in the BM. For instance, one of the best results is achieved by 100 MW LIB integration in the Southern GSP group which is equal to £10.68m or £26m when there is a CAPEX decrease of 30% or 70%, respectively.

Our results demonstrate that PSH offers significant benefits, particularly in terms of monthly BM cost reduction (~£373k for the 100 MW case). The study also found that LIB offered similar potential reductions in BM costs and emissions to PSH. For 100 MW (2-hour duration) monthly cost reduction of ~£349k and emission reduction of ~216 kgCO2/h is possible. Due to its high round trip efficiency, LIB is effective even in regions with relatively stable generation patterns and limited intermittency, making it a reasonable choice in many GSP groups. Based on this, it can be said that PSH and LIB demonstrate more cost and emission reduction potential across both seasons compared to others. The results indicate that VRFB, while not as effective as LIB in some GSP groups (due to the lower round-trip efficiency), can still contribute to cost reduction in specific regions, especially during the summer months. However, it shows a limited impact on emission reduction with 100 MW integration. It can be concluded that VRFBs lag LIBs in terms of effectiveness and efficiency but still show potential if costs decrease. For HES, our results show that its use is highly region-specific, with certain GSP groups responding more favourably during both winter and summer seasons (e.g. monthly summer BM cost reduction will be by ~£292k for 100 MW case). HES may not significantly reduce costs in some regions but may offer reasonable emission reduction in others (e.g. monthly winter emission reduction will be by ~97 kgCO2/h for 1 MW case).

**Impact of storage location:** The paper provides a detailed evaluation of the impact network location has on how energy storage system would operate in the BM, and how this impacts balancing costs and emissions. This information helps in understanding the practical implications of adopting these technologies in the context of grid balancing. An important finding is that storage location makes a significant difference to balancing cost reduction and carbon emissions reduction, and in many locations only one of these is improved. This shows the importance of accounting for network location and the details of electricity market mechanisms when making energy storage investment planning decisions. In addition, it motivates the need for BM market reforms. Based on the findings, the only GSP group that provides both monthly BM cost reduction and carbon emission reduction is the Northern GSP group with 100 MW LIB, where both balancing cost and emission in the summer month are reduced by ~£151k and ~135 kgCO2/h simultaneously. This shows the particular potential of this GSP group as a suitable location for the integration of this storage technology. By focusing on the variability between regions in terms of cost and carbon reduction, for 100 MW integration, in the winter, the cost reduction variability ranges from £4,630.14 for HES to £64,408.41 for PSH, while change in carbon emission reduction variability ranges from +344.43 kgCO2/h for HES to +747.92 kgCO2/h for PSH. In the summer, the cost variability increases significantly, with LIB showing a range from £31,251.29 (East Midlands) to £349,311.47 (Southern), and PSH showing a range from £93,902.23 (East Midlands) to £372,761.49 (Southern) while change in carbon emission variability ranges from +299.42 kgCO2/h for PSH to +1,288.69 kgCO2/h for HES. This shows how the location is important and also this underscores the importance of seasonal factors to optimize both economic and environmental outcomes. This suggests that policymakers might consider introducing locational market reforms to better manage the variability in the GB energy system.

**Carbon emissions Impact:** The study also assessed the effects of energy storage system integration on carbon emissions and generation output. It highlights the potential environmental benefits of reducing carbon emissions through the use of cleaner energy sources and efficient storage system. Regarding the marginal changes in carbon emissions during the winter months, we observed that VRFB, and HES exhibited the most number of GSP group with reduction followed by PSH and then LIB. Most of variability is for PSH (+221 to -98 kgCO2/h), following closely by LIB (+230 to -71 kgCO2/h). During the summer months, this situation is different and LIB and PSH have the most number of GSP groups with emission reduction and the most variability is for HES (+20 to -18 kgCO2/h), followed by LIB (+8 to -27 kgCO2/h). When it comes to the impact of grid-scale energy storage system integration on emissions during the winter season, a small number of regions stood out as the most effective locations for energy storage. The variability of emission differences for PSH is the highest (+853 to +105 kgCO2/h) and all the GSP groups show emission increase. In the summer, LIB generated emission reductions in five regions and the best value at all is for LIB integration within the Northern GSP group with emission reduction value of 134.74 (kgCO2/h) as a notable result and in this situation, highest variability is for HES (+1288 to -14 kgCO2/h). These findings underscore the significant potential for emissions reduction through the deployment of LIB, particularly grid-scale systems and specifically within some limited regions. At the same time, the results show the inadequacy of incentives for carbon emissions reduction. We find that there is currently a lack of incentives for storage developers to locate projects where they would have the greatest impact on carbon emissions, highlighting the need for new market mechanisms on the path to net-zero, such as stronger carbon prices or the integration of explicit carbon emissions targets into the BM dispatch process.

## METHODS

### Balancing mechanism modelling

In this paper, the aim is creating a high-fidelity model for the GB network with detailed recent data that accurately replicates the functioning of the BM. The model developed to simulate the BM activities including storage is described in this section. The BM is a pay-as-bid market based on half-hour rolling auctions, with gate closure one hour before real-time. Every half hour, the BM indicates the cost of providing power during that period.[25] Generators and suppliers are required to publish their intended generation or demand details for each settlement period, known as Physical Notifications. At gate closure, these Physical Notifications become finalized and are referred to as Final Physical Notifications (FPNs). During the auction window, market participants submit "bids" or "offers" into the BM. A bid is the price they propose to either consume more electricity or generate less (downward), while an offer is the price to consume less or generate more electricity (upward). In this process, the system operator accepts bids and offers from BM units to change their power injection, which can be done by producers changing their generation and consumers changing their demand. The offer cost is that the system operator must remunerate unit *u* for decreasing consumption and the bid cost is that unit *u* has to compensate the system operator, in the event of increasing the consumption or decreasing the generation. So, the bid cost is subtracted from the offer cost in each time period. The objective function minimizing the overall BM cost including the marginal degradation cost of the energy storage systems is given in (1).

$$\text{Min} \quad \sum_{t \in T} \left( \sum_{u \in U} \left( C_{t,u}^{\text{offer}} - C_{t,u}^{\text{bid}} \right) + \sum_{s \in S} C_{t,s}^{\text{deg}} \right) \quad (1)$$

where, *t* represents the time and *s* denotes the storage unit. The objective function is subject to constraints such as bounds on the power output of BM units, power balance, power flow, voltage magnitude and angle limits, offer and bid orders, and constraints related to the storage systems' dynamics. The detailed mathematical modelling developed for the BM is reported in the supplemental information.

### Great Britain's power grid and data

The power grid in GB supplies electricity throughout England, Scotland, and Wales. It is a complex system that comprises power stations, high-voltage transmission lines, and substations (see Figure S1(A)). The network is divided into 14 grid supply point (GSP) groups which are also called distribution network operator (DNO) areas (see Figure S1(B)) (The GSP group IDs and related region names are listed in Table S1)[26]. NGESO is responsible for managing and operating the power grid in GB and ensures the balance between electricity supply and demand in real-time, maintaining grid stability and reliability. The electricity generation mix in GB has been evolving in recent years and has been transitioning towards a cleaner and more renewable energy mix, reducing its reliance on coal-fired power plants. It includes a combination of different energy sources, such as nuclear power, natural gas, wind power, solar power, biomass, and hydroelectricity. Besides, GB is connected to neighbouring power grids through interconnectors. These interconnectors facilitate the import and export of electricity between countries.[27]

To account for seasonality, we have used data covering both January and July of 2022. The collection of bids and offer orders is performed via Elexon's Balancing and Settlement code.[25] Through Elexon's SAA-I014 raw settlement files[28], we gathered the FPNs and energy imbalances (i.e. differences between the scheduled power and the actual metered one). The GB transmission network is adapted from the comprehensive National Grid Electricity Ten-Year Statement (NG ETYS). Figure S1 demonstrates an overview of the grid, comprising 1982 nodes ranging from 33 kV to 400 kV, including AC lines and transformers, along with HVDC links. The model implementation involved Python 3.8 and Pyomo[29], and is solved using CPLEX 22.1.1[30] on a Core i7 CPU with 32 GB of RAM. The average computation time for each day, consisting of 48 half-hour settlement periods, is about 25 minutes. To validate the proposed approach, the BM costs are estimated by our model for the time periods considered in January and July of 2022 and the results are £191.86m and £263.16m, respectively, while this cost for the mentioned months by NGESO calculations[31] is £173.85m and £240.10m, showing that our model reasonably approximate BM costs. Additional modelling details, further parameters calculation, and a detailed justification are reported in the supplemental information.

### Carbon emission study

One of the important contributions of this paper is the estimate of the carbon emission reduction due to the usage of energy storage systems. To calculate this, the power differences for each generation unit between a test case with energy storage system integration and the base case without energy storage system is multiplied by the carbon intensity of that generation technology and summed to obtain an estimation of emission change. The marginal carbon intensity used for different types of generation units is listed in Table 1.[32-34]

Table 1: Marginal carbon emission intensity of different power generation technologies

| Power generation technology | Emission (kgCO2/MWh) |
|---|---|
| Coal | 937 |
| Open Cycle Gas Turbine (OCGT) | 651 |
| Combined Cycle Gas Turbine (CCGT) | 394 |
| Other | 300 |
| Biomass | 120 |
| Nuclear | 0 |
| Non-pumped storage hydropower (NPSHYD) | 0 |
| Wind | 0 |
| Pumped storage hydropower (PSH) | 0 |
| ESS/DSR/Embedded Renewables | 0 |

### Energy storage technologies

In this paper, we undertake a thorough comparative study involving four different energy storage systems: lithium-ion battery (LIB), vanadium redox flow battery (VRFB), pumped storage hydropower (PSH), and hydrogen energy storage (HES). The LIB, VRFB, and PSH have been selected based on their widespread use and established track records worldwide. HES is included as a comparison considering that it is expected to be a critical technology in the future. Important parameters related to these technologies have been obtained from[35-39] and the details are as follows:

**Power capacity:** Our study encompasses two power capacities, evaluating the impact of the energy storage systems at both 1 MW and 100 MW levels. The 1 MW case has been included to assess the marginal impact of deploying energy storage systems. Note that the 1 MW case for PSH and HES is only realistic in the context of considering a marginal change to the size of a larger system during design, and thus the investment cost parameters have been obtained by proportionally scaling those for the 100 MW case i.e. the cost values for 100 MW cases have been divided by 100 to be considered as the 1 MW cases costs. This selection allows us to explore their scalability and adaptability for marginal and grid-scale study conditions.

**Charging and discharging efficiency:** Efficiency is a critical parameter in evaluating energy storage systems. We have taken into account the round-trip efficiencies, which are set at 85% for LIB, 64% for VRFB, 79% for PSH, and 30% for HES.[40] These values directly impact the overall effectiveness and energy conservation potential of each technology.

**Degradation costs:** Degradation costs play a pivotal role in understanding the long-term economic viability of energy storage systems. For this, we are using a linear model where degradation cost is proportional to the energy storage throughput, and it is also applied to both charging and discharging. We have assigned degradation costs of 13.17 £/MWh for LIB,[38] 0.78 £/MWh for VRFB,[35] 0.00 £/MWh for PSH,[41] and 0.23 £/MWh for HES,[42] reflecting the impact of wear and tear over time The main parts of PSH are reservoir, powerhouse, and electromechanical elements that with proper maintenance, suitable performance can be sustained. For HES, the costly components are the fuel cell and electrolyser. HES degradation is based on these elements' depreciation.

This selection of technologies, capacities, efficiencies, and degradation costs allows us to conduct a comparative study. Our analysis aims to provide insights into the performance, feasibility, and economic aspects of these energy storage systems, enabling us to make informed decisions about their suitability for BM application within the GB's power grid.

Net Present Value calculation

Our analysis includes evaluation of the cost-effectiveness of the energy storage systems, considering their upfront, operational and maintenance expenditures using Net Present Value (NPV) assessment. Additionally, the NPV values enable us to assess the financial viability of employing energy storage systems. Here, in addition to cash flow from the energy storage systems participation in BM, revenues from value stacking (i.e. a strategy where storage systems can derive revenue from multiple sources and this maximizes the financial return by participating in various markets and services simultaneously) through participation in the Capacity Market (CM)[43] and provision of Dynamic Containment (DC) service[44] is also considered. The Capacity Market (CM) provides extra capacity payments on top of BM participation. It ensures that there is sufficient capacity to meet peak electricity demand, rewarding participants for being available to supply power or reduce demand when needed. The Dynamic Containment (DC) service, on the other hand, is a fast-acting frequency response service that helps maintain grid stability by quickly adjusting power output in response to frequency deviations. In our value stacking approach, we allocate excess capacity after BM participation to the DC service to raise the overall revenue which the capacity for BM participation is based on the assumption of perfect forecasting. The required data related to the CM and DC service have been extracted from EnAppSys[45]. For the full lifetime of each type of storage technology, the NPV can be defined as:

$$\begin{aligned}
\mathbf{NPV} &= \sum_{y=0}^{Y} \frac{Cash\,Flow_y}{(1+d)^y} \\
&= \sum_{y=0}^{Y} \left( \frac{k.NBB_y + CMR_y + (1-k).DCR_y - C_{O\&M,y}^{fix}}{(1+d)^y} \right) - CAPEX + \left( \frac{RV}{(1+d)^{Y+1}} \right)
\end{aligned} \quad (2)$$

where, *NBB* is the net BM benefit including the profit (in terms of BM cost reduction) gained through energy storage system integration compared with the base case without the storage, while accounting for degradation and charging costs. *CMR* and *DCR* are respectively the revenue stream of energy storage system via Capacity Market and Dynamic Containment service. $\kappa$ is a fraction specifying how energy storage system capacity is split between the BM and DC services. For this, after extracting the storage capacity percentage participation in BM, the remaining would be the capacity percentage that can provide DC service which varies between 39% to 54% depending on technology, location, and season. For the CM participation a de-rating factor of 20%[46] has been employed. $C_{O\&M,y}^{fix}$ is the operation and maintenance fixed cost in each year. These values are summarized in Table 2.[7,47,48] It should be clarified that to have comparable settings for all the energy storage systems, *CAPEX* values considered for PSH and HES are the ones for 4-hours duration from[40] divided by 2. The discount rate is considered by *d* and equals to 8% for all technologies. The *CAPEX* is the fixed investment cost, and the *RV* is the end-of-life residual value[49] computed as a fraction of the initial investment cost as:

$$RV = \lambda_{RV} \times CAPEX \quad (3)$$

where, the considered value of factor $\lambda_{RV}$ is 20%[50], 40%[51], 20%[52], and 5%[53] for LIB, VRFB, PSH, and HES, respectively.

**Table 2: Economic parameters for different energy storage systems**

| Technology | LIB | | VRFB | |
|---|---|---|---|---|
| Power Capacity (MW) | 1 | 100 | 1 | 100 |
| Duration (Hour) | 2 | 2 | 2 | 2 |
| CAPEX (£) | 539,348.37 | 41,895,060.23 | 753,820.03 | 60,452,744.77 |
| Fixed O&M Cost (£) | 2,563 | 206,969 | 4,527 | 358,962 |
| Operational Life (Year) | 11 | 11 | 12 | 12 |
| **Technology** | **PSH** | | **HES** | |
| Power Capacity (MW) | 1 | 100 | 1 | 100 |
| Duration (Hour) | 2 | 2 | 2 | 2 |
| CAPEX (£) | 912,765.79 | 91,276,578.54 | 2,344,571.11 | 234,457,110.52 |
| Fixed O&M Cost (£) | 6,468 | 646,778 | 18,595 | 1,859,487 |
| Operational Life (Year) | 60 | 60 | 30 | 30 |

## SUPPLEMENTAL INFORMATION

Supplemental information can be found online at supplemental information.

## ACKNOWLEDGMENTS

The authors acknowledge the Engineering and Physical Sciences Research Council (EPSRC) for grant funding under reference number EP/W027321/1.

## EXPERIMENTAL PROCEDURES

### Resource Availability

#### Lead Contact

Further information and requests for resources should be directed to and will be fulfilled by the Lead Contact, Seyyed Mostafa Nosratabadi (mostafa.nosratabadi@eng.ox.ac.uk, sm.nosratabadi@yahoo.com).

#### Data and Code Availability

Original data for this paper, used for simulation and analyses, are archived at a public repository and are available on GitHub: https://github.com/EsaLaboratory/DIGEST

## AUTHOR CONTRIBUTIONS

Conceptualization, S.M.N., I.S. and T.M.; Methodology, S.M.N., I.S. and T.M.; Software, S.M.N., I.S.; Validation, S.M.N., I.S., V.K., P.G., M.A., D.A.H. and T.M.; Investigation, S.M.N., I.S. and T.M.; Writing – Original Draft, S.M.N.; Writing –Review & Editing, S.M.N., I.S., V.K., P.G., M.A., D.A.H. and T.M.; Resources, S.M.N., I.S., V.K., P.G., M.A., D.A.H. and T.M.; Project Administration, T.M.

## DECLARATION OF INTERESTS

The authors declare no competing interests.